\documentclass[traditabstract, longauth]{aa}
\usepackage{natbib, graphicx,amssymb, url}
\begin{document}

\title{Differential Frequency-dependent Delay from the Pulsar Magnetosphere}
\author{
T.~E.~Hassall\inst{ \ref{soton}} \and
B.~W.~Stappers\inst{ \ref{jod}} \and
P.~Weltevrede\inst{ \ref{jod}} \and
J.~W.~T.~Hessels\inst{ \ref{astron} \and \ref{uva}} \and
A.~Alexov\inst{ \ref{uva} \and \ref{stsci}} \and
T.~Coenen\inst{ \ref{uva}} \and
A.~Karastergiou\inst{ \ref{ox}} \and
M.~Kramer\inst{ \ref{mpifr} \and \ref{jod}} \and
E.~F.~Keane\inst{ \ref{mpifr}} \and
V.~I.~Kondratiev\inst{ \ref{astron}} \and
J.~van~Leeuwen\inst{ \ref{astron} \and \ref{uva}} \and
A.~Noutsos\inst{ \ref{mpifr}} \and
M.~Pilia\inst{ \ref{astron}} \and
M.~Serylak\inst{ \ref{nancay} \and \ref{cnrs}} \and
C.~Sobey\inst{ \ref{mpifr}} \and
K.~Zagkouris\inst{ \ref{ox}} \and
R.~Fender\inst{ \ref{soton}} \and
M.~E.~Bell\inst{ \ref{caastro} \and \ref{soton}} \and
J.~Broderick\inst{ \ref{soton}} \and
J.~Eisl\"offel\inst{ \ref{tls}} \and
H.~Falcke\inst{ \ref{nijmegen} \and \ref{astron}} \and
J.-M.~Grie\ss{}meier\inst{  \ref{cnrs} \and \ref{nancay}} \and
M.~Kuniyoshi\inst{ \ref{mpifr}} \and
J.~C.~A.~Miller-Jones\inst{ \ref{curtin} \and \ref{uva}} \and
M.~W.~Wise\inst{ \ref{astron} \and \ref{uva}} \and
O.~Wucknitz\inst{ \ref{mpifr} \and \ref{ubonn}} \and
P.~Zarka\inst{ \ref{meudon}} \and
A.~Asgekar\inst{ \ref{astron}} \and
F.~Batejat\inst{ \ref{oso}} \and
M.~J.~Bentum\inst{ \ref{astron}} \and
G.~Bernardi\inst{ \ref{kapteyn}} \and
P.~Best\inst{ \ref{roe}} \and
A.~Bonafede\inst{ \ref{hamburg} \and \ref{bremen}} \and
F.~Breitling\inst{ \ref{aip}} \and
M.~Br\"uggen\inst{ \ref{hamburg}} \and
H.~R.~Butcher\inst{ \ref{astron} \and \ref{anu}} \and
B.~Ciardi\inst{ \ref{mpifa}} \and
F.~de~Gasperin\inst{ \ref{hamburg} \and \ref{mpifa}} \and
J.-P.~de~Reijer\inst{ \ref{astron}} \and
S.~Duscha\inst{ \ref{astron}} \and
R.~A.~Fallows\inst{ \ref{astron}} \and
C.~Ferrari\inst{ \ref{nice}} \and
W.~Frieswijk\inst{ \ref{astron}} \and
M.~A.~Garrett\inst{ \ref{astron} \and \ref{leiden}} \and
A.~W.~Gunst\inst{ \ref{astron}} \and
G.~Heald\inst{ \ref{astron}} \and
M.~Hoeft\inst{ \ref{tls}} \and
E.~Juette\inst{ \ref{raiub}} \and
P.~Maat\inst{ \ref{astron}} \and
J.~P.~McKean\inst{ \ref{astron}} \and
M.~J.~Norden\inst{ \ref{astron}} \and
M.~Pandey-Pommier\inst{ \ref{leiden} \and \ref{lyon}} \and
R.~Pizzo\inst{ \ref{astron}} \and
A.~G.~Polatidis\inst{ \ref{astron}} \and
W.~Reich\inst{ \ref{mpifr}} \and
H.~R\"ottgering\inst{ \ref{leiden}} \and
J.~Sluman\inst{ \ref{astron}} \and
Y.~Tang\inst{ \ref{astron}} \and
C.~Tasse\inst{ \ref{meudon}} \and
R.~Vermeulen\inst{ \ref{astron}} \and
R.~J.~van~Weeren\inst{ \ref{cfa} \and \ref{leiden} \and \ref{astron}} \and
S.~J.~Wijnholds\inst{ \ref{astron}} \and
S.~Yatawatta\inst{ \ref{astron}} 
}

\institute{
School of Physics and Astronomy, University of Southampton, Southampton, SO17 1BJ, UK \label{soton} \and
Jodrell Bank Center for Astrophysics, School of Physics and Astronomy, The University of Manchester, Manchester M13 9PL,UK \label{jod} \and
Netherlands Institute for Radio Astronomy (ASTRON), Postbus 2, 7990 AA Dwingeloo, The Netherlands \label{astron} \and
Astronomical Institute 'Anton Pannekoek', University of Amsterdam, Postbus 94249, 1090 GE Amsterdam, The Netherlands \label{uva} \and
Space Telescope Science Institute, 3700 San Martin Drive, Baltimore, MD 21218, USA \label{stsci} \and
Astrophysics, University of Oxford, Denys Wilkinson Building, Keble Road, Oxford OX1 3RH \label{ox} \and
Max-Planck-Institut f\"{u}r Radioastronomie, Auf dem H\"ugel 69, 53121 Bonn, Germany \label{mpifr} \and
Station de Radioastronomie de Nan\c{c}ay, Observatoire de Paris, CNRS/INSU, 18330 Nan\c{c}ay, France \label{nancay} \and
Laboratoire de Physique et Chimie de l' Environnement et de l' Espace, LPC2E UMR 7328 CNRS, 45071 Orl\'{e}ans Cedex 02, France \label{cnrs} \and
ARC Centre of Excellence for All-sky astrophysics (CAASTRO), Sydney Institute of Astronomy, University of Sydney Australia \label{caastro} \and
Th\"{u}ringer Landessternwarte, Sternwarte 5, D-07778 Tautenburg, Germany \label{tls} \and
Department of Astrophysics/IMAPP, Radboud University Nijmegen, P.O. Box 9010, 6500 GL Nijmegen, The Netherlands \label{nijmegen} \and
International Centre for Radio Astronomy Research - Curtin University, GPO Box U1987, Perth, WA 6845, Australia \label{curtin} \and
Argelander-Institut f\"{u}r Astronomie, University of Bonn, Auf dem H\"{u}gel 71, 53121, Bonn, Germany \label{ubonn} \and
LESIA, UMR CNRS 8109, Observatoire de Paris, 92195 Meudon, France \label{meudon} \and
Onsala Space Observatory, Dept. of Earth and Space Sciences, Chalmers University of Technology, SE-43992 Onsala, Sweden \label{oso} \and
Kapteyn Astronomical Institute, PO Box 800, 9700 AV Groningen, The Netherlands \label{kapteyn} \and
Institute for Astronomy, University of Edinburgh, Royal Observatory of Edinburgh, Blackford Hill, Edinburgh EH9 3HJ, UK \label{roe} \and
University of Hamburg, Gojenbergsweg 112, 21029 Hamburg, Germany \label{hamburg} \and
Jacobs University Bremen, Campus Ring 1, 28759 Bremen, Germany \label{bremen} \and
Leibniz-Institut f\"{u}r Astrophysik Potsdam (AIP), An der Sternwarte 16, 14482 Potsdam, Germany \label{aip} \and
Research School of Astronomy and Astrophysics, Australian National University, Mt Stromlo Obs., via Cotter Road, Weston, A.C.T. 2611, Australia \label{anu} \and
Max Planck Institute for Astrophysics, Karl Schwarzschild Str. 1, 85741 Garching, Germany \label{mpifa} \and
Laboratoire Lagrange, UMR7293, Universit\`{e} de Nice Sophia-Antipolis, CNRS, Observatoire de la C\'{o}te d'Azur, 06300 Nice, France \label{nice} \and
Leiden Observatory, Leiden University, PO Box 9513, 2300 RA Leiden, The Netherlands \label{leiden} \and
Astronomisches Institut der Ruhr-Universit\"{a}t Bochum, Universitaetsstrasse 150, 44780 Bochum, Germany \label{raiub} \and
Centre de Recherche Astrophysique de Lyon, Observatoire de Lyon, 9 av Charles Andr\'{e}, 69561 Saint Genis Laval Cedex, France \label{lyon} \and
Harvard-Smithsonian Center for Astrophysics, 60 Garden Street, Cambridge, MA 02138, USA \label{cfa} 
}

\abstract{
Some radio pulsars show clear `drifting subpulses', in which subpulses are seen to drift in pulse longitude in a systematic pattern. Here we examine how the drifting subpulses of PSR B0809+74 evolve with time and observing frequency. We show that the subpulse period ($P_3$) is constant on timescales of days, months and years, and between $14 - 5100$~MHz. Despite this, the shapes of the driftbands change radically with frequency. Previous studies have concluded that, while the subpulses appear to move through the pulse window approximately linearly at low frequencies ($< 500$~MHz), a discrete step of $\sim180^\circ$ in subpulse phase is observed at higher frequencies ($>820$~MHz) near to the peak of the average pulse profile. We use LOFAR, GMRT, GBT, WSRT and Effelsberg 100-m data to explore the frequency-dependence of this phase step. We show that the size of the subpulse phase step increases gradually, and is observable even at low frequencies. We attribute the subpulse phase step to the presence of two separate driftbands, whose relative arrival times vary with frequency -- one driftband arriving 30 pulses earlier at 20~MHz than it does at 1380~MHz, whilst the other arrives simultaneously at all frequencies. The drifting pattern which is observed here cannot be explained by either the rotating carousel model or the surface oscillation model, and could provide new insight into the physical processes happening within the pulsar magnetosphere.
}

\keywords{pulsars: general -- pulsars: individual: PSR B0809+74 -- telescopes: LOFAR}
\authorrunning{Hassall et al.}
\titlerunning{Frequency-dependent Delay from the Pulsar Magnetosphere}
\maketitle

\section{Introduction}
PSR B0809+74 is a nearby \citep[0.43~kpc,][]{bbgt02} pulsar, whose proximity to Earth and relative brightness mean that it can be clearly detected at all observing frequencies between $\sim12$~MHz and $\sim10$~GHz \citep[see, e.g.][]{brc75, bup+86, pw86, sgg+95, es03, rd11, gjk12}. The broadband detectability of the source, as well as the fact that the pulsar exhibits many interesting features such as `drifting subpulses' \citep{dc68}, `nulling' \citep{bac70} and `microstructure' \citep{ccd68}, makes it an excellent source for studying the pulsar magnetosphere and the pulsar emission mechanism. Here we focus on PSR B0809+74's drifting subpulses.

Drifting subpulses were first identified by \cite{dc68}, who noticed that the single-pulse components (`subpulses') of some pulsars are not randomly distributed across the emission region. Instead, they appear to drift through the pulse window in a well-defined pattern. The drift rate of subpulses is normally characterised by three parameters \citep[as defined by][]{sspw70}:
$P_1$, the pulsar's rotation period;
$P_2$, the interval between subpulses; and 
$P_3$, the separation between two driftbands. 

Initially, drifting subpulses were attributed to oscillations of the neutron star surface. This idea was  proposed with the discovery of drifting subpulses \citep{dc68} and was further developed by \cite{van80}, who showed that  the surface of a neutron star can support $p$-mode oscillations, vibrations caused by pressure waves. These surface oscillations modulate the pulsed emission, and if the waves are in a beat frequency with the rotation period then the oscillations are observed as drifting subpulses.

Surface oscillation models were widely abandoned in favour of the so-called `carousel' model \citep{rs75}, because of their failure in explaining several features in the observed drifting patterns - these include the conservation of pulse longitude across nulls \citep{urwe78}, the high degree of stability of the drift bands \citep{seps70}, and the changing drift-rate after a null \citep{la83}. In the carousel model, radio emission is generated from discrete locations (`sparks') in the plasma of the pulsar magnetosphere. The configuration of the sparks with respect to each other remains fixed, but the whole sparking region rotates around the magnetic pole, like a carousel. This rotation (caused by $\mathbf{ E \times B}$ drift) moves the sparks slightly between pulses, so the subpulses appear to drift. The rate of rotation has been measured for a few pulsars, allowing detailed maps of the polar cap to be made \citep[see for example,][]{dr01}, and modified versions of this model have been used to try to explain all of the features listed above \citep[see e.g.][]{fr82, gs00,gmg03, ghm+08}.

\begin{table*}
\centering
\caption{Summary of the observations presented here.}
\label{tab:B0809+74_obs}
\begin{tabular}{rcccccc}
\hline
Observatory	& Centre  & Bandwidth & Sampling & $N_\mathrm{pulses}$ & MJD & Reference\\
			&Frequency (MHz)	& (MHz) & Time (ms)  &	&	&	\\
\hline
LOFAR $^a$	& 38				&	48		&	1.3103	& 5571 	& 55815	&This work\\
LOFAR $^b$	& 160			&	9		&	1.3103	& 178283 	& 54896	&This work\\
LOFAR $^c$	& 145			&	12		&	1.3103	& 185628 	& 55099	&This work\\
LOFAR $^d$	& 143			&	48		&	0.6554	& 33433 	& 55822	&This work\\
LOFAR $^e$	& 209			&	38		&	0.8192	& 5572  	& 56008	&This work\\
WSRT 		& 328 			&	10		&	0.8192	& 15984 	& 51874	&\cite{es03}\\
GMRT 		& 624			&	20		&      0.2288      & 5696     & 55609	&\cite{gjk12}\\
GBT 			& 820 			&	200		&	0.1600	& 478    	& 54944	&\cite{rd11}\\
WSRT 		& 1380 			&	80		&	0.4096	& 13092 	& 52694	&\cite{es03}\\
WSRT 		& 2220 			&	120		&	0.8192	& 2788  	& 56021	&This work\\
Effelsberg 	& 4850			&	500		&	0.1000	& 4100  	& 51487	&This work\\
\hline
\end{tabular}
\tablefoot{
LOFAR observation IDs:
\tablefoottext{a}{L30803}
\tablefoottext{b}{L2009\_11193}
\tablefoottext{c}{L2009\_14591}
\tablefoottext{d}{L30910}
\tablefoottext{e}{L53897}
}
\end{table*}

More recently however, it has been shown that drifting subpulses can indeed be explained by surface oscillations, as long as the oscillations are centred on the magnetic axis, and not the rotation axis \citep{cr04,cr08}. These `non-radial' oscillations address many of the problems mentioned earlier, and it is now possible to produce quantitative non-radial oscillation models of known drifting pulsars \citep[including PSR B0809+74 and PSR B0943+10,][]{rd11, rc08}, which fit the data at least as well as the carousel model. 

A particularly interesting feature of PSR B0809+74's driftbands is the step in subpulse phase (i.e. phase in the $P_3$ direction) which is detected at observing frequencies of 820~MHz and above \citep{es03, rd11}, but is unseen at lower frequencies \citep{brc75, pw86, es03}. One of the strengths of the surface oscillation model presented in \cite{rd11} is that it can be used to explain the subpulse phase step which appears at high frequencies, by invoking the existence of a nodal line, which moves into the line-of-sight. In this paper we use radio observations from $14-5100$~MHz to examine the driftbands of PSR B0809+74 in detail and map their evolution as a function of frequency, focussing in particular on the subpulse phase step. These new observations are used to test current models of the pulsar magnetosphere.

\section{Observations}
\label{sec:obs}

Our analysis is based on newly acquired LOFAR observations, spanning the lowest 4 octaves of the observable `radio  window', as well as archival data from several other radio telescopes. LOFAR has two sets of antennas which observe in separate frequency bands -- the Low Band Antennas (LBAs) and the High Band Antennas (HBAs). The LBAs were used to observe the pulsar from $14-62$~MHz. The pulsar was also observed using the HBAs from $119-167$~MHz and from $190-228$~MHz. All of the LOFAR observations in this paper were taken using the coherent sum of the six stations on the `Superterp' (central core of the array), except the HBA data from MJD 54896 (which were taken using the 4-tile LOFAR test station) and MJD 55099 (which were taken using the incoherent sum of 3 remote stations). For further information on LOFAR's beamformed observing modes see \cite{sha+11}, and for a general LOFAR description see van Haarlem et al. (submitted).

In addition to the LOFAR observations detailed above, we also present the results of our analysis on data at 328~MHz and 1380~MHz from the Westerbork Synthesis Radio Telescope \citep[WSRT,][]{es03},  624~MHz data from the Giant Meterwave Radio Telescope \citep[GMRT,][]{gjk12}, 820~MHz data from the Green Bank Telescope \citep[GBT,][]{rd11}, previously unpublished 4850~MHz data from the Effelsberg telescope \citep[see][for details of the system]{ljk+08} and new 2220~MHz data from WSRT \citep[see][for details of the system]{kss08}. Full details of all of the observations are summarised in Table~\ref{tab:B0809+74_obs}.

\section{Analysis}
\label{sec:anal}
\subsection{Initial Processing}
Data from each observation were de-dispersed to a DM of 5.75 pc~cm$^{-3}$ and collapsed in frequency so that each observation (with the exception of the one taken with the LOFAR LBAs) contained a single frequency channel. At low frequencies (below $\sim80$~MHz) the pulse profile of PSR B0809+74 changes significantly over the wide fractional bandwidth of the LOFAR LBAs \citep[see, e.g.][]{hsh+12}, so the observation was divided into 4$\times$12~MHz bands, with centre frequencies of 20~MHz, 32~MHz, 44~MHz and 56~MHz. Initial estimates of the pulsar's period at the time of each observation were derived from known spin parameters \citep{hlk+04}. These estimates were optimised by performing a narrow search in period space around the ephemeris values. Each optimised period was used to divide the data from the appropriate observation into pulse-period-sized segments and create a `pulse stack', a two-dimensional array of the intensity of the source as a function of pulse longitude and pulse number. We used these pulse stacks to produce fluctuation spectra, subpulse phase tracks, and $P_3$-folded pulse stacks, in order to examine the temporal and spectral stability of the drift patterns, and test models of the pulsar magnetosphere.

\subsection{Fluctuation Spectra}
To investigate the periodicity associated with drifting subpulses, we used the Longitude-Resolved Fluctuation Spectrum \citep[LRFS, see][]{bac70}. In the LRFS method, data are first divided into blocks of successive pulses (typically 512 pulses per block). For each block, Fourier transforms are calculated for constant pulse longitude columns. The fluctuation spectra produced from all of the blocks are combined to produce an average fluctuation spectrum at each pulse longitude. If the pulsar has drifting subpulses, each pulse longitude bin which features a driftband will have a frequency associated with the distance between subsequent driftbands. These lead to `features' in the LRFS, which can be used to determine $P_3$. 

\subsection{Subpulse Phase Tracks}
\begin{figure}
\begin{center}
\includegraphics[width=\linewidth]{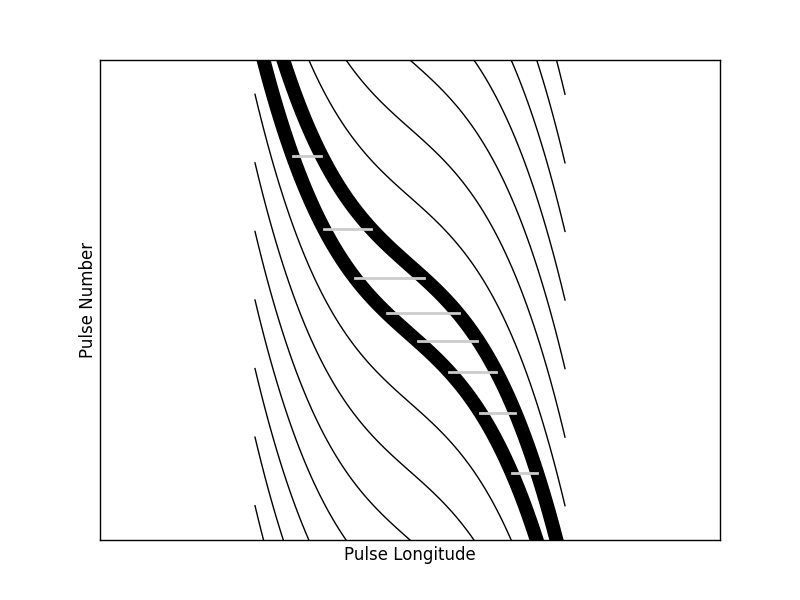}
\caption{Illustration of how curved driftbands mean that $P_2$ varies with pulse longitude. The black lines represent the shape of simulated driftbands. The grey lines represent the size of $P_2$ for the highlighted driftbands. One can see that $P_2$ varies significantly depending on the gradient of the driftbands at a given pulse longitude.}
\label{fig:p2_phase_dep}
\end{center}
\end{figure}

For most pulsars, $P_2$ is not constant. Driftbands are often curved, and as a consequence, $P_2$ varies significantly with pulse longitude \citep{es02}. Figure~\ref{fig:p2_phase_dep} illustrates this -- the black lines represent curved driftbands in a pulse stack, and in the highlighted driftbands, $P_2$ is shown at various pulse longitudes by the grey lines. The curvature of the driftbands produces different values of $P_2$ at different pulse longitudes. Techniques like the Two-dimensional Fluctuation Spectrum \citep[2DFS, see][]{es02} cannot resolve any pulse longitude-dependence, and effectively provide the $P_2$ distribution averaged over the whole pulse window. To investigate the shape of the driftbands, we used the complex spectra produced from the LRFS. By calculating the complex phases in each pulse longitude bin it is possible to produce a subpulse phase track, which is a measure of the average shape of a driftband. For a more detailed explanation of this process see \cite{es02}, and for a description of the specific method used in this paper see \cite{wwj12}\footnote{ Note that in this paper, we define subpulse phase with the opposite sign to \cite{es03} and other authors. We find this more intuitive, as it means that the subpulse phase tracks have a positive slope for subpulses which drift in the positive direction, so the two can be compared directly.}.

\subsection{$P_3$ Folding}
Subpulse phase tracks are good for probing the stability of driftbands from observation-to-observation, and also provide good indicators of the shape of the driftband when the signal-to-noise ratio of the observations are low. However, in some circumstances they do not give the whole picture. The subpulse phase track only gives a single value for the subpulse phase at each pulse longitude bin, so if there are two overlapping driftbands present in the same pulse longitude bin \citep[as is suggested to be the case in PSR B0809+74, see][]{es03} the technique will not accurately describe the data.

An alternative way of analysing the data, which can help with this problem, is $P_3$ folding. This is done by folding the pulse stack (which has already been folded at the rotation period of the pulsar) at the $P_3$ value derived from the LRFS, to produce a 2-D array containing the shape of the `average' driftband.  Similar methods have been used in the past by \cite{dr01}, \cite{vkr+02} and \cite{bmr11}. PSR B0809+74 is a nulling pulsar and when the pulsar goes into its null state, its drifting pattern is disrupted and the driftbands quickly go out of phase with each other, causing the folded driftbands to become smeared together. In addition, even without nulls the value of $P_3$ is not stable from pulse-to-pulse and will constantly fluctuate around a mean value. To mitigate these effects, we folded the data in blocks of 256 pulses, which were added together using an iterative technique to maximise the correlation between subsequent blocks of subbands, by allowing arbitrary phase offsets between blocks.  Analysis through the $P_3$-folding technique is complementary to analysis with subpulse phase tracks, as $P_3$-folded data provides extra information on where the power is in the pulse profile and can be interpreted more intuitively, but the technique does not work as well on low signal-to-noise data.

\section{Results}
\label{sec:results}

\subsection{Temporal Stability of Driftbands}
\label{sec:temp}

\begin{figure*}
\begin{center}
\includegraphics[width=\textwidth , clip=true]{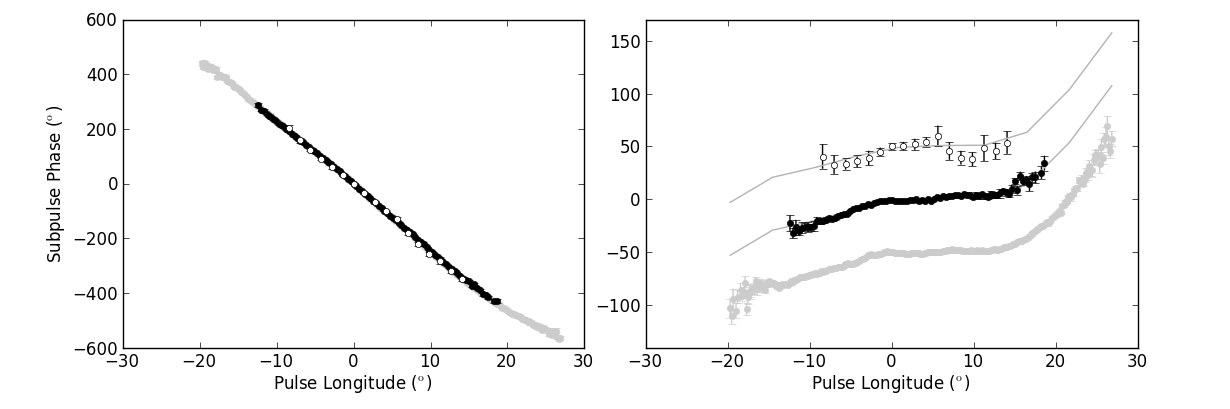}
\caption{The subpulse phase tracks of three LOFAR observations of PSR B0809+74 taken in March 2009 (white points), September 2009 (black points) and September 2011 (grey points).  The right panel shows the same tracks with a slope of $-25^{\circ / \circ}$ subtracted and offset by $\pm50^\circ$, so that fine structure is easier to see. The grey line is a spline of the most recent observation so that data can be easily compared across epochs. The shape of the subpulse phase tracks are constant to within experimental uncertainties.}
\label{fig:temp_long}
\end{center}
\end{figure*}

\begin{figure*}
\begin{center}
\includegraphics[width=\textwidth , clip=true]{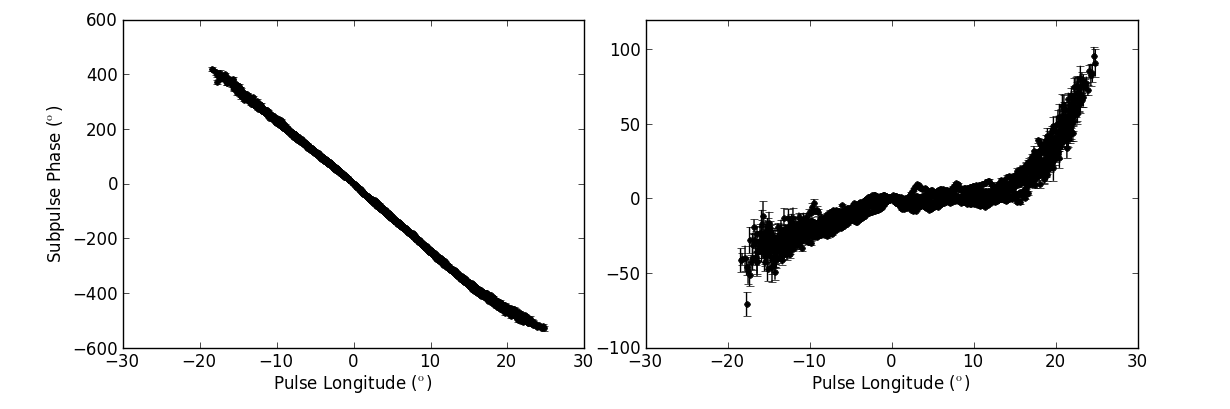}
\caption{The subpulse phase tracks of 8$\times$1.5~hour blocks of a LOFAR HBA observation of PSR B0809+74, showing that the shape subpulse phase track remains approximately constant with time (typical spread is $\sim10^\circ$). The right panel shows the same tracks with a slope of $-25^{\circ / \circ}$ subtracted, so that fine structure is easier to see.}
\label{fig:temp_short}
\end{center}
\end{figure*}

\cite{la83} showed that the drift-rate of PSR B0809+74 changes regularly on very short timescales, increasing slightly after nulls, before relaxing to its pre-null speed after $\sim20$~pulses. The pulsar's drift-rate has also been shown to vary on timescales of several minutes by \cite{vkr+02}, who observed a `slow' mode which was stable for $\sim120$~pulses\footnote{We note that this slow mode must be quite rare and/or frequency dependent, as it was not observed in any of the long observations presented here.}. To investigate the temporal stability of the drift pattern of PSR B0809+74 on longer timescales, we used three LOFAR HBA observations spanning a period of 2 years. Using the LRFS, we found $P_3$ to be constant across all observations to within experimental uncertainties. 

The curved shape of the driftband means that $P_2$ varies between 30~ms (just before the subpulse phase step) and 43~ms (at the subpulse phase step) within a single cycle of subpulse phase, however the shape of the subpulse phase tracks remains stable on very long timescales. The subpulse phase tracks of the three observations are shown in the left panel of Figure~\ref{fig:temp_long}. The right panel of the figure shows the subpulse phase tracks with a constant gradient of $-25$ degrees (of subpulse phase) per degree (of pulse longitude) subtracted to show the fine structure of the subbands in more detail. This slope of $-25^{\circ / \circ}$ was chosen because it was the best straight-line fit to the data. From the figure, one can see that the subpulse phase tracks are identical (to within experimental uncertainties) in observations taken in March 2009 (white points) September 2009 (black points) and September 2011 (grey points). The only noticeable difference between the tracks is their extent in pulse longitude -- the most recent observations have more data points at the edge of the pulse profile compared with those taken previously. However, this is simply a reflection of LOFAR's available collecting area and bandwidth increasing\footnote{The observations from March 2009 used only 4 HBA tiles, the September 2009 observation used the incoherent sum of 3 stations (each with 48 HBA tiles) and the September 2011 observation was taken using the coherent sum of the Superterp (288 HBA tiles total).}. The fact that there are no intrinsic changes to the subpulse phase tracks shows that the subpulses are stable on timescales of months to years. 

We also used the most recent observation to investigate subpulse stability on shorter timescales. We broke the 2011 LOFAR HBA observation into 8 pieces (each of 4096 pulses), and produced subpulse phase tracks for each piece. These are plotted in Figure~\ref{fig:temp_short}. As in Figure~\ref{fig:temp_long}, the right panel in the figure shows the same subpulse phase tracks with a constant slope of $-25^{\circ / \circ}$ subtracted. Again, the subpulse phase tracks look stable, although the spread on these points is noticeably larger than the spread of the long timescale phase tracks. We attribute the increased spread to pulse-to-pulse jitter. \cite{hmt75} claimed that the average pulse profile of PSR B0809+74 becomes stable after $\sim300$ pulses, although \cite{lkl+12} have recently shown that the stabilisation timescales for pulsars are actually much longer ($\gtrsim10^4$ pulses) than expected. Driftbands may also need many pulses before they stabilise. This could explain the difference in the size of the subpulse phase steps measured at 820~MHz by \cite{rd11}, who used relatively short integration times of 15~minutes ($<500$ pulses) in their observations. Despite the pulse-to-pulse jitter, we note that the spread in the subpulse phase tracks is still small -- typically of order $\sim10^\circ$. Thus subpulse phase tracks seem to be stable on all of the timescales we have probed (i.e. hours, days, months and years).

\subsection{Spectral Stability of Driftbands}
We now compare the drifting behaviour of the pulsar at different frequencies. From the preceding discussion, we know that the driftbands are stable on timescales of hours, days, months and years, so -- although most of the observations we are using are not simultaneous -- we can confidently assume that any differences observed with frequency are not caused by temporal variations. 

\subsubsection{Subpulse Phase Tracks}
It has been shown previously that the shape of PSR B0809+74's driftbands vary significantly as a function of frequency. Above 820~MHz, \citep{wbs81, pw86, es03, rd11} there is a sudden step of more than 100$^\circ$ in subpulse phase near to the peak of the pulse profile. But below 500~MHz, several authors (\citealp{es03} at 328~MHz, \citealp{pw86} at 400~MHz and \citealp{brc75} at 500~MHz) have searched for similar discontinuities and failed to find any evidence for a subpulse phase step. 

\begin{figure}
\centering
\includegraphics[height=0.9\textheight]{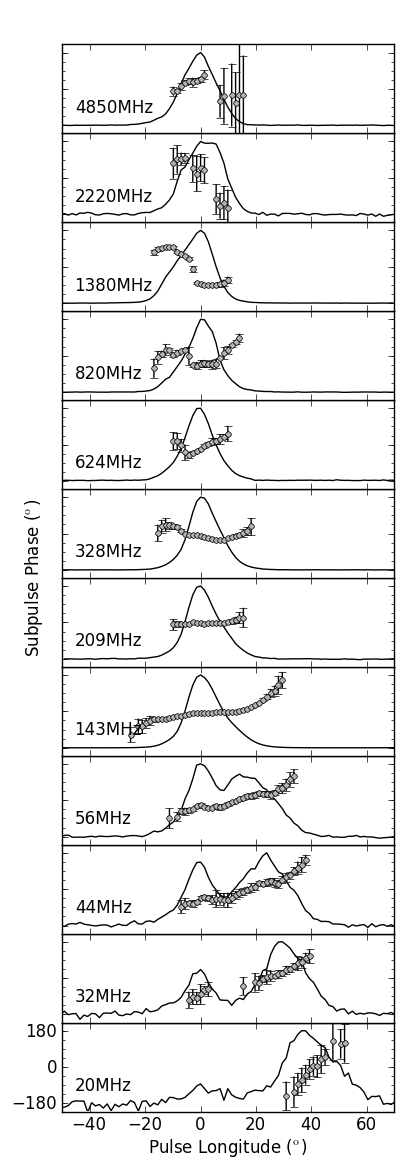}
\caption[The subpulse phase tracks of PSR B0809+74 between $14-5100$~MHz.]{The subpulse phase tracks (grey points) and average pulse profiles (black lines) of PSR B0809+74 between $14-5100$~MHz. All phase tracks have had a constant slope of $-25^{\circ / \circ}$ removed. The central observing frequency of each panel is indicated in the bottom left corner.}
\label{fig:0809_tracks}
\end{figure}

Figure~\ref{fig:0809_tracks} shows the pulse profiles and subpulse phase tracks obtained from the analysis of observations between 14~MHz and 5100~MHz. In \cite{hsh+12}, we showed that one of the components in PSR B0809+74's average pulse profile (the component on the leading edge at 624~MHz and below, and the trailing edge at 820~MHz and above) remains fixed in pulse longitude at the fiducial point, and the second component moves through it as a function of frequency. The pulse profiles in Figure~\ref{fig:0809_tracks} have been aligned so that this `fiducial component' is at $0^\circ$ pulse longitude.  As in Section~\ref{sec:temp}, the subpulse phase tracks have all had a gradient of $-25^{\circ / \circ}$ removed from them, to accentuate fine structure. The subpulse phase step is clearly visible at 1380~MHz, but what is apparent is that it gradually increases in size as a function of frequency. It is visible at frequencies as low as 328~MHz, where it can be seen on the leading edge of the pulse profile. Even at 143~MHz, the subpulse phase track is asymmetric, and the leading edge looks like it has been disrupted. The subpulse phase step moves to later pulse longitudes and increases in magnitude as observing frequency increases, reaching its maximum size of $\sim190^\circ$ at 2220~MHz. The size of the phase step is slightly lower in the 4850~MHz data, although it is possible that it may have wrapped in subpulse phase, i.e. because the driftbands recur periodically, the size of the phase step could be wrong by $\pm360^\circ$. Despite the large variation in the shape of the driftbands, we find (using the LFRS) that $P_3$ is constant at all frequencies.

\begin{figure*}
 \centering
 \begin{tabular}{cccc}
\includegraphics[width=0.23\textwidth]{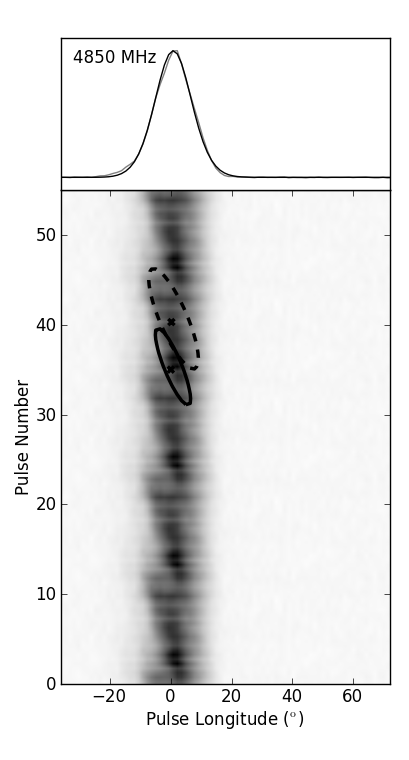} &\includegraphics[width=0.23\textwidth]{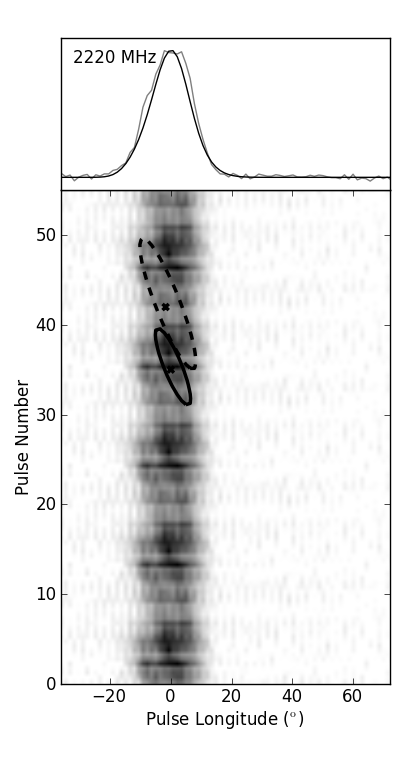} &\includegraphics[width=0.23\textwidth]{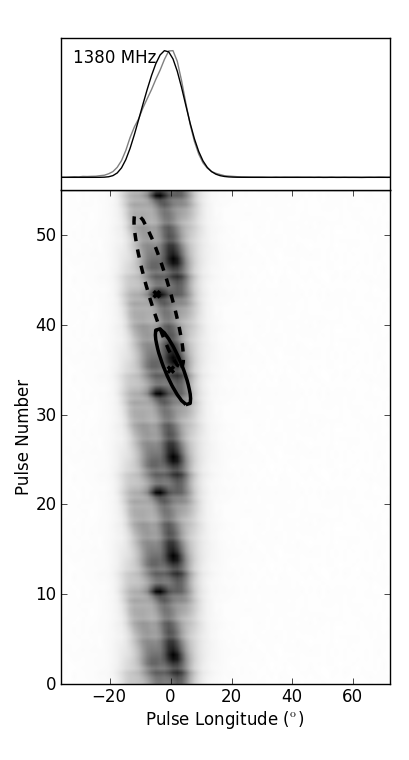} &\includegraphics[width=0.23\textwidth]{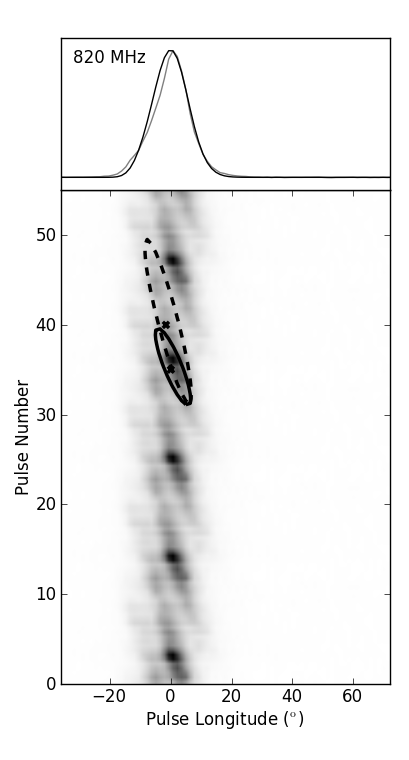} \\ 
\includegraphics[width=0.23\textwidth]{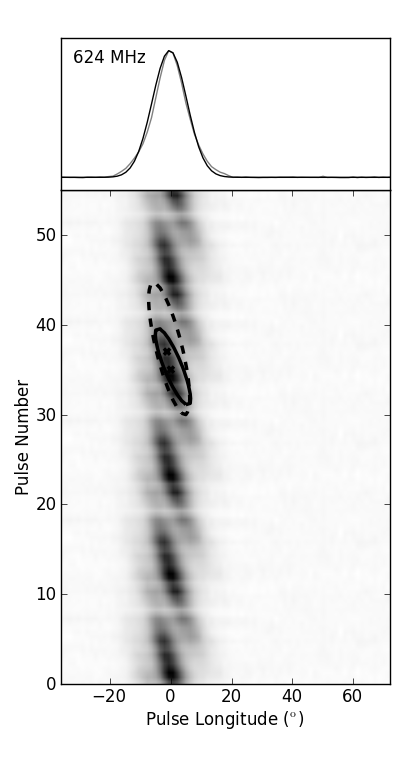} & \includegraphics[width=0.23\textwidth]{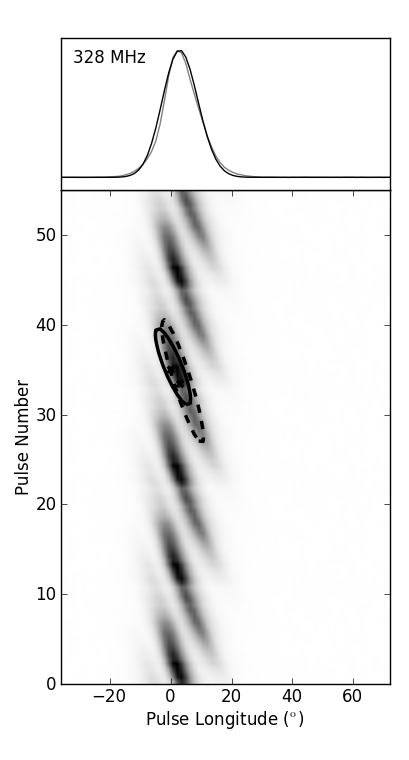} &\includegraphics[width=0.23\textwidth]{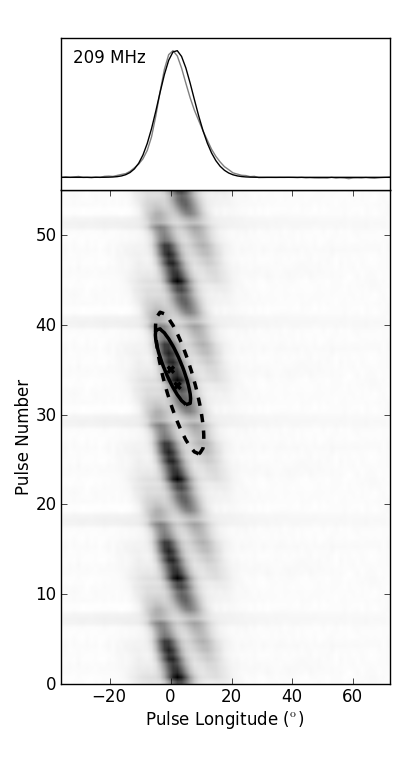} & \includegraphics[width=0.23\textwidth]{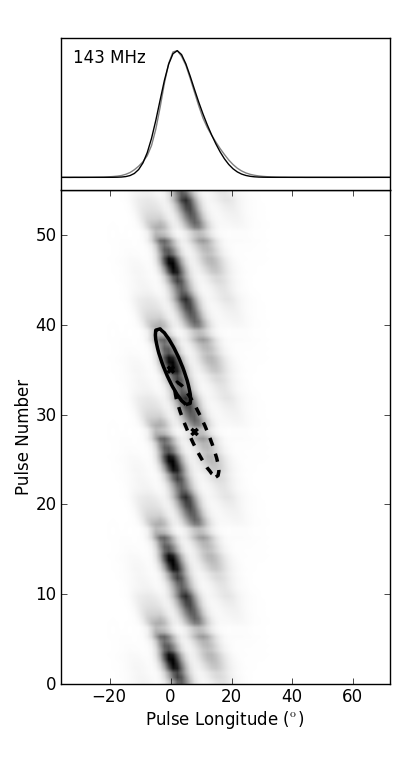} \\ 
\includegraphics[width=0.23\textwidth]{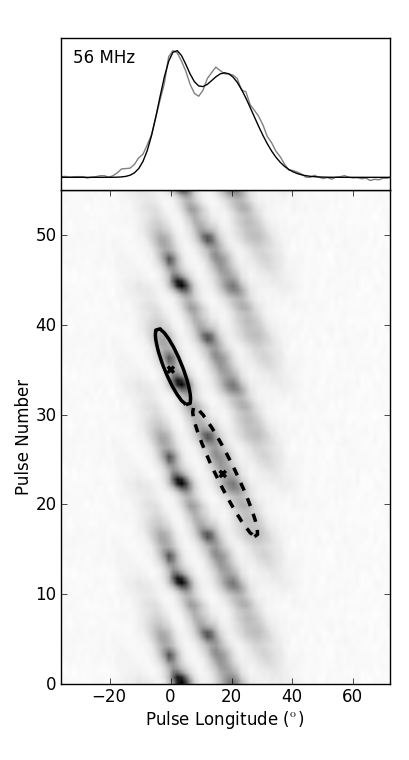} & \includegraphics[width=0.23\textwidth]{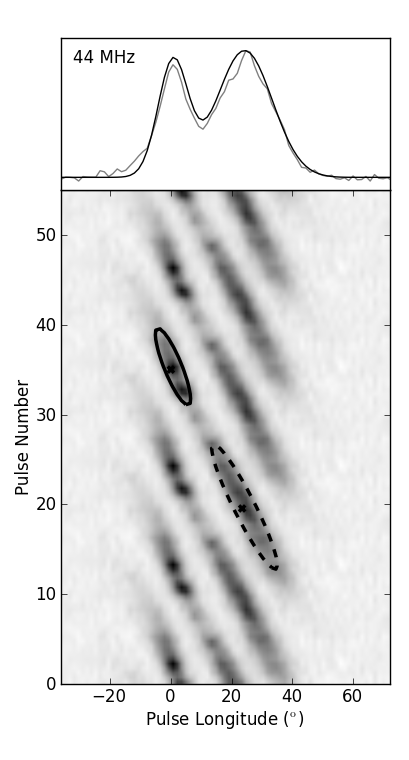} & \includegraphics[width=0.23\textwidth]{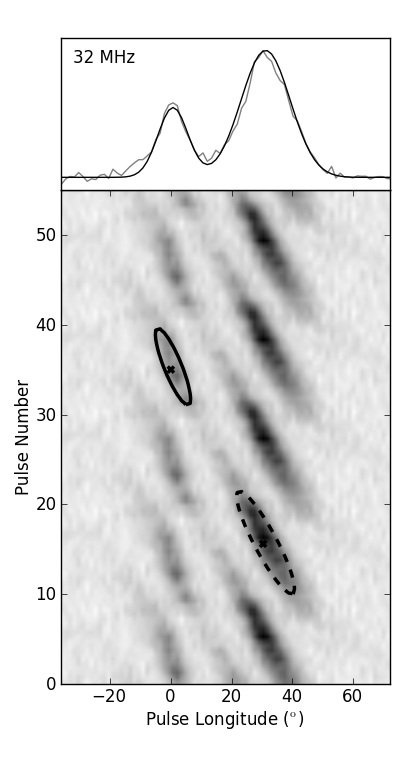} & \includegraphics[width=0.23\textwidth]{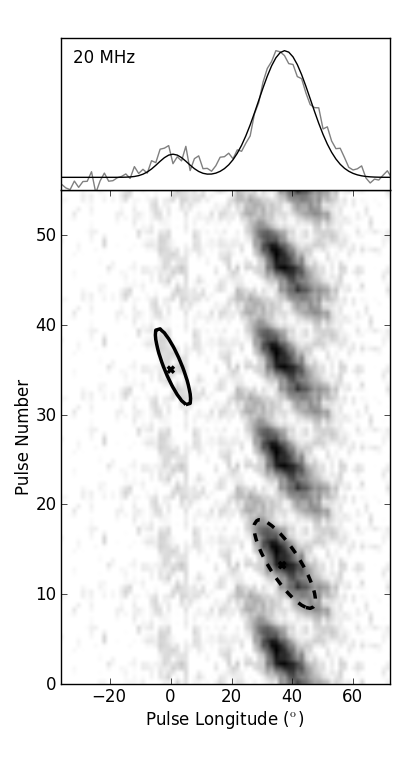} \\ 
 \end{tabular}
 \caption{Folded `pulse stacks' from each of the observations used. The solid and dashed ellipses show the half power points of 2D gaussians fitted to the data using a least squares method. The central observing frequency of each panel is indicated in the top left corner. The driftbands are repeated over $5\times P_3$ so that the full drift pattern is visible at all frequencies. The pulse profile (grey line), and a simulated pulse profile (black line) from each 2D fit is shown in the panel above the corresponding pulse stack.}
 \label{fig:subpulses_freq}
 \end{figure*}

\subsubsection{$P_3$ Folds}
The $P_3$-folded pulse stacks of each of our observations are shown in Figure~\ref{fig:subpulses_freq}. The pulse stacks look radically different at each frequency. There are two distinct driftbands at the lowest frequencies, which are seemingly associated with the two components in the pulse profile, and the driftbands move closer together in both pulse longitude and subpulse phase with increasing frequency. This movement is most prominent in the LOFAR LBA observations, where the    centroid of the rightmost driftband appears to arrive 10~pulses later at 56~MHz than it does at 20~MHz, and $23.5^\mathrm{o}$ earlier in pulse longitude. These observations (at 20, 32, 44 and 56~MHz) were taken simultaneously, and the plots shown in Figure~\ref{fig:subpulses_freq} are time aligned. 

The fact that one driftband remains stationary, whilst the other suffers a frequency-dependent delay of 10 pulses is contrary to the expectations of radius-to-frequency mapping. Although one could de-disperse the data to a different DM to make it appear that the two components move away from each other at the same rate in the pulse profile, the DM of the pulsar is not large enough to account for the 10-pulse delay in the $P_3$ direction, which only affects one  driftband. Thus we conclude that there are two sets of driftbands -- the `fiducial driftband'  (the driftband associated with the fiducial point of the pulse profile) appears fixed in both pulse longitude and subpulse phase \citep[in strong agreement with the results of][]{hsh+12}, and the other component suffers a frequency-dependent delay relative to it.
 
At higher frequencies, the driftbands begin to overlap, and above 143~MHz it is hard to ascertain which driftband is which by eye. To determine what happens above 143~MHz, we fitted both of the folded driftbands with 2D gaussians using a least squares method. To alleviate some of the degeneracy in the region where the two driftbands overlap, the shape of the fiducial driftband was held constant for the fits\footnote{We note that in the LBA observations, where the two driftbands are distinct, the shape of the fiducial driftband is approximately constant.}. The half-power contours for the fitted gaussians are shown in Figure~\ref{fig:subpulses_freq}. The solid line represents the 2D gaussian fitted to the fiducial driftband, and the dashed line represents the gaussian fitted to the other driftband. In the figure, the folded pulse stacks above 56~MHz have been aligned such that the fiducial driftband appears at a fixed pulse longitude and subpulse phase. The fits reveal that above 56~MHz second driftband continues to move through the pulse profile in both pulse longitude and subpulse phase. The moving driftband at 20~MHz arrives 30 pulses earlier than at 1380~MHz, whilst emission from the other driftband arrives at the same pulse longitude and subpulse phase at all frequencies. We note that in the regions where the two driftbands overlap, they are not aligned. This suggests that the two driftbands are not part of the same system, and that the delay we see is not simply a result of a pulse window function. We attribute the fact that the driftbands appear to be aligned at 56~MHz to coincidence -- the movement of the non-fiducial driftband spans $\sim40$ pulses over the frequency range observed here, so it is not unreasonable that the driftbands should appear to be aligned over a narrow frequency range. The subpulse phase step, which appears when the driftbands begin to overlap also suggests that the two driftbands are not perfectly aligned.

\begin{figure}
\centering
\includegraphics[width=\linewidth]{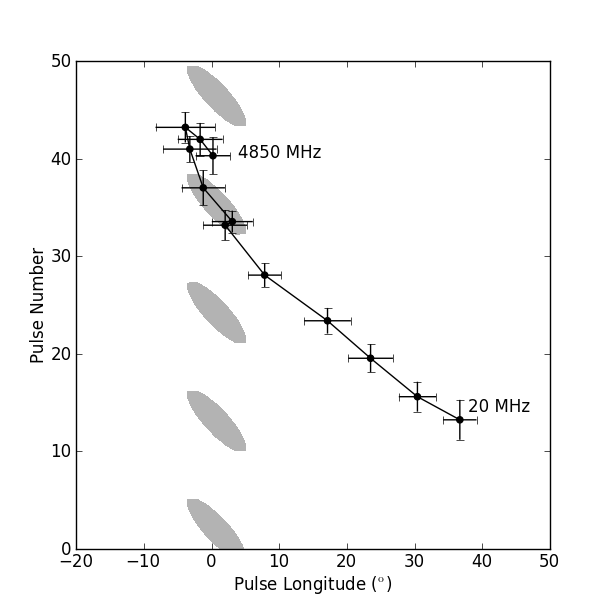}
\caption{The path of the moving driftband as a function of frequency. The shape of the stationary (`fiducial') driftband corresponding to the fiducial pulse profile component is shown in grey.}
\label{fig:path}
\end{figure}

Figure~\ref{fig:path} shows the path of the moving driftband as a function of frequency. The black points show the position of the centroid of the moving driftband and the grey regions each represent the position and shape of the fiducial component's driftband. One can see that the driftband follows a smooth, and continuous path through the figure. Above 1380~MHz the driftband appears to move back towards later pulse longitude and earlier pulse number, although it should be noted that this is the region in our data where the signal-to-noise is relatively poor and the two sets of driftbands are overlapping, so there is an increased degeneracy in the fits. 

\section{Discussion}
\label{sec:disc}
In Section~\ref{sec:results}, we showed that the radio emission from PSR B0809+74 comes from two separate driftband structures. One driftband remains fixed at the fiducial point found in \cite{hsh+12} and in subpulse phase, whilst the other driftband moves towards earlier pulse longitude and later pulse numbers with increasing frequency. The movement of the driftband is subtle above 143~MHz, and it only moves significantly in the LOFAR LBA band, where, until now, narrow available bandwidths have prevented this effect from being seen. 

Above 143~MHz, the two driftbands overlap with each other in pulse longitude, and comparing the $P_3$-folded pulse stacks in Figure~\ref{fig:subpulses_freq} with the position of the corresponding subpulse phase step in Figure~\ref{fig:0809_tracks}, it is clear that the interface between the two different driftbands is what is producing the subpulse phase step.  This driftband configuration replaces the need for the `absorption feature' \citep{bar81, ran83} on the leading edge of the pulse profile, and could potentially explain the orthogonal polarisation modes seen in PSR B0809+74's single pulses. Certainly, the fits to the $P_3$-folded pulse stacks shown here look very similar to the islands of polarisation seen in Figures~5 and 6 of \cite{rrs05}, although unfortunately we do not yet have polarisation data to test this hypothesis directly.  In the following discussion, we attempt to reconcile our findings with models of the pulsar magnetosphere.

\subsection{Surface Oscillation Model}
Currently, the surface oscillation model cannot explain many of the phenomena shown here. One of the requirements of the surface oscillation model of \cite{cr04} is that nodal lines should introduce subpulse phase steps of exactly $180^\mathrm{\circ}$.  The authors explained a $120^\mathrm{\circ}$ subpulse phase step seen by \cite{es03} in 1380~MHz PSR B0809+74 data by suggesting that because the emission we see is an average over a finite frequency range, the phase step is ``washed out''. This is a valid argument for why one frequency would have a phase step not exactly equal to $180^\mathrm{\circ}$, but it is unlikely that the subpulse phase step would increase so systematically with frequency (see Figure~\ref{fig:0809_tracks}), despite the differing bandwidths in each observation, if this were the case for all frequencies.

A second requirement of the model is that the spacing of the components should follow the same distribution as a spherical harmonic sampled along a single line-of-sight. We can rule this out qualitatively for PSR B0809+74 by noting that the size of the fiducial component is approximately constant at all frequencies, and the spacing between the components in pulse longitude varies (smoothly) from $\sim-5^\mathrm{\circ}$ to $\sim+35^\mathrm{\circ}$. The model also cannot explain how the subpulse phase step begins on the leading edge of the pulse profile at low frequencies and moves through the central component, appearing on the trailing edge of the component at high frequencies.

\subsection{Carousel Model}
The behaviour seen here also cannot currently be explained using the carousel model.  Using a carousel, it is impossible to recreate the two components which are distinct below 143~MHz and overlap at higher frequencies. The carousel model has an inherent symmetry to it, and when components in the pulse profile move apart they should do so symmetrically to fit with the hollow cone described by the inner edges of the dipolar magnetic field. The average pulse profile of PSR B0809+74 (particularly its spectral evolution) is very asymmetric. Most scenarios which are able to explain the movement of the component require one half of the carousel to be missing. Absorption has been used in the past to explain this asymmetry \citep{bar81, ran83}, but to account for what we observe here, absorption needs to selectively effect only the leading component at low frequencies, and the trailing component at high frequencies, which seems unlikely. Even using a distorted polar cap \citep[as discussed in][]{ae98}, it is impossible to reproduce the pulsar's profile evolution. In the latest $\mathbf{ E \times B}$ models \citep[e.g.][]{ta12}, the currents and potential differences in the magnetosphere can be asymmetric along the line of sight. \cite{vt12} used this fact to explain the two different drift speeds seen in PSR B0826-34. Despite this, such models do not change the geometry of the system, and therefore cannot explain the two sets of driftbands, or the movement of one driftband through the other as a function of frequency.

Even invoking two carousels (one for each driftband), the emission from one driftband must undergo a strongly frequency-dependent delay in subpulse phase whilst all of the emission from the other must arrive almost simultaneously\footnote{Note that this argument is equally valid for rejecting a model with two surface-oscillation modes.}. If we take this delay at face value, as a physical 30-pulse delay, the size of the region needed to produce such a delay by light-travel time is $\>10^7$~km, or $\sim180$ times the radius of the light cylinder. Alternatively, we can interpret the changing drift pattern as being due to the one pulse-profile component moving to later pulse longitudes, being modulated by an underlying driftband. This scenario still requires a delay of $40^\mathrm{\circ}$ in pulse longitude ($\sim0.14$~seconds), corresponding to a light-travel time of 43000~km, roughly 70\% of the radius of the light cylinder. Both interpretations are in stark contrast with the results of \cite{hsh+12}, where we showed that all of the emission from the fiducial component in the pulse profile between 46~MHz and 8~GHz must originate from within a region smaller than 400~km. 

\subsection{Birefringence}
In \cite{hsh+12}, we suggested that birefringence could be responsible for the evolution of the pulse profile with frequency, but the fact that the shape and pitch angle of the moving driftband changes with frequency suggests that it is unlikely to be caused by a simple, axis-symmetric propagation effect. If it were, the driftbands would move relative to each other but the shape and pitch angle of the driftbands would remain constant at all frequencies. Perhaps refraction could account for the observed frequency-dependence when coupled with an asymmetric plasma distribution in the magnetosphere, although the frequency-dependent delay also seems large for a propagation effect which must occur in a region smaller than the light cylinder ($10^5$~km).

\section{Conclusion}
We have shown that both the shape, and the separation between the driftbands ($P_3$) of PSR B0809+74 are stable on timescales of $\sim$years. But, whilst $P_3$ does not change, the shape of the driftbands is extremely variable with frequency. Emission from PSR B0809+74 is composed of two separate driftbands. The emission from one of the driftbands arrives at the same pulse longitude in the same pulse at all frequencies \citep[the fiducial point found in][]{hsh+12}, whilst emission from the other driftband changes location in pulse longitude and is delayed by tens of pulses in subpulse phase at high frequencies. The subpulse phase step, which is visible at high frequencies, is attributed to the interface of the two driftbands \citep[as suggested by][]{es03}. 

The carousel and the surface-oscillation models are not emission mechanisms per se, and so, should not be expected to explain frequency-dependent effects. However, the features seen in the spectral evolution of the driftbands shown in our observations seem incompatible with current incarnations of either model. We are, as of yet, unable to find any emission mechanism or geometry to explain these features satisfactorily. Similar studies on the spectral evolution of subpulse phase in other pulsars at low frequencies will be done to determine whether PSR B0809+74 is a rare case, or if this type of frequency evolution is common to the pulsar population. More data could also provide further clues to the mechanism causing the effect. In particular, low-frequency polarisation data would be very useful in determining whether this phenomenon is related to the orthogonal polarisation modes, as seen by \cite{rrs05} and others. It could also be used to extract more information about the conditions of the magnetosphere in the regions where each of the driftbands originated.

\section*{Acknowledgements}
The authors would like to thank Axel Jessner, Rachel Rosen, Paul Demorest, Vishal Gajjar and Balchandra Joshi for sharing data with us, and Geoff Wright for his insight and helpful discussions. 

TEH is funded by European Research Council Advanced Grant 267697 ``4 Pi Sky: Extreme Astrophysics with Revolutionary Radio Telescopes''. BWS and PW are supported through an STFC rolling grant. JWTH is a Veni Fellow of the Netherlands Foundation for Scientific Research. JvL and TC are supported by the Netherlands Research School for Astronomy (Grant NOVA3- NW3-2.3.1) and by the European Commission (Grant FP7- PEOPLE-2007-4-3-IRG \#224838). AK is grateful to the Leverhulme Trust for financial support. CS is supported by the DFG (German Research Foundation) within the framework of the Research Unit FOR 1254, Magnetisation of Interstellar and Intergalactic Media: The Prospects of Low-Frequency Radio Observations. CF acknowledges financial support by the {\it ÒAgence Nationale de la RechercheÓ} through grant ANR-09-JCJC-0001-01.

LOFAR, the Low Frequency Array designed and constructed by ASTRON, has facilities in several countries, that are owned by various parties (each with their own funding sources), and that are collectively operated by the International LOFAR Telescope (ILT) foundation under a joint scientific policy. This publication made use of observations taken with the 100-m telescope of the MPIfR (Max-Planck-Institut f\"ur Radioastronomie) at Effelsberg. The Westerbork Synthesis Radio Telescope is operated by the ASTRON (Netherlands Foundation for Research in Astronomy) with support from the Netherlands Foundation for Scientific Research NWO. The National Radio Astronomy Observatory is a facility of the National Science Foundation operated under cooperative agreement by Associated Universities, Inc. This research has made use of NASA's Astrophysics Data System Bibliographic Services.

\bibliography{refs}
\bibliographystyle{aa}
\end{document}